\documentclass{llncs}
\usepackage{graphicx,amssymb}
\usepackage{makeidx}
\usepackage{moreverb}

\title{Formal study of plane Delaunay triangulation}
\author{Jean-Fran\c{c}ois Dufourd\inst{1} \and Yves Bertot\inst{2}
\thanks{This work is supported by the French ANR project GALAPAGOS (2007-2010).}
}

\institute{Universit\'{e} de Strasbourg \\ 
LSIIT, UMR CNRS-UdS 7005,\\
P\^{o}le API,
Boulevard S. Brant, BP 10413, 67412 Illkirch, France
\email{dufourd@lsiit.u-strasbg.fr}
\and INRIA-Centre de Sophia Antipolis M\'editerran\'ee,\\ 
2004, Route des Lucioles, 06902 Sophia-Antipolis Cedex, France\\
\email{Yves.Bertot@sophia.inria.fr}}

\begin{document}

\maketitle  
 
\begin{abstract}
  This article presents the formal proof of correctness for a plane
  Delaunay triangulation algorithm.  It consists in repeating a
  sequence of edge flippings from an initial triangulation until the
  Delaunay property is achieved.  To describe triangulations, we rely
  on a combinatorial hypermap specification framework we have been
  developing for years.  We embed hypermaps in the plane by attaching
  coordinates to elements in a consistent way.  We then describe what
  are legal and illegal Delaunay edges and a flipping operation which
  we show preserves hypermap, triangulation, and embedding invariants.
  To prove the termination of the algorithm, we use a generic approach
  expressing that any non-cyclic relation is well-founded when working
  on a finite set.
\end{abstract}

\bibliographystyle{plain}

\section{Introduction}
Delaunay triangulation is one of the cornerstones of computational
geometry.  In two dimensions, the task is, given a collection of input
points, to find triangles whose corners are the input points, so that
none of the input points lies inside the circumcircle of a triangle.
This constraint about circumcircles makes it possible to ensure that
flat triangles are avoided as much as possible.  This is important for
many numeric simulation applications, as flatter triangles imply more
errors in the simulation process.

To our knowledge, this article presents the first formalized proof of
correctness of an algorithm to build a plane Delaunay
triangulation. The algorithm takes as input an arbitrary triangulation
and repeatedly flips illegal edges until the Delaunay criterion is
achieved.  This is one of the most naive algorithms, but proving its
formal correctness is already a challenge.    We shall review more
related work around geometry, combinatorial maps, and formalization
in section~\ref{RW}.

We use a general data-structure to represent plane subdivisions and
perform proofs, known as hypermaps
 \cite{tut2,cor,duf09b,ber:duf,duf:pui,duf08a,duf09}.
Hypermaps are collections of {\em darts} equipped with two permutations.
Darts are elementary objects, more elementary than points: usually,
two darts constitute an edge and several darts constitutes a point.
The two permutations are used to describe how darts are connected
together to constitute an edge or a point.
We then need to give locations to points.  This is done by {\em
  embedding} the darts in the plane, by simply attaching coordinates,
making sure that all darts that constitute the
same point should have the same coordinates.  We then restrict our
work to {\em triangulations} by defining a way to compute faces
and by considering hypermaps with three-point faces.
  We shall review our approach to hypermaps
in section~\ref{HY}.

The edge flipping operation can be defined at a topological level: it
mainly consists in detaching an edge from two points and attaching it
back to two other points.  As an intermediate step, we observe a
hypermap that is not a triangulation, but after re-attaching the edge
we get back a new triangulation.  We review the topological aspect of
edge flipping in section~\ref{FL}.

The next step is to describe where edge flipping occurs.  At this point
the coordinates of points play a role.  We formalize how oriented
triangles and circumcircles are computed and define {\em illegal edges}.
We show that illegal edges can be flipped and that the operation also
preserves the geometric constraints of well-embedded triangulations.  We
study this aspect in section~\ref{DP}.

A crucial aspect of our formalization is to show that the algorithm
terminates.  We tackle this issue by formalizing the argument that the
number of possible triangulations based on a given collection of darts
and a given collection of points is finite.  We then exhibit a real
number associated to each triangulation that decreases when an illegal
edge is flipped.  Because the
set of possible triangulations is finite, this is enough to ensure
termination.  This point is studied in a generic manner in section~\ref{TF}.

In section~\ref{DS}, we show the kind of correctness statement that we have
proved about our Delaunay algorithm.  The full formalization is
developed in Coq \cite{ber:cas,Coq}.  It covers many different
aspects: hypermaps, geometry, termination problems.  Because of the
 size of this
paper, we do not enter into details, but the full
formalization is available at \cite{duf:ber10}.

\section{Related work} 
\label{RW}
 
\subsection{Geometric modeling and Delaunay triangulations}
Like
\cite{gs85}, we work with a general model of plane subdivisions, based
on hypermaps 
\cite{cor} and combinatorial oriented maps \cite{tut2}.  The triangulations
of our development are a kind of combinatorial oriented maps.

Triangulations are widely used in computational geometry to model,
reconstruct or visualize surfaces. For instance, the CGAL library
offers a lot of advanced functionalities about triangulations
\cite{boi:yvi02}. Among them, the Delaunay triangulations
\cite{gs85,knu,ede00,deb08} are very
appreciated in applications because their triangles are regular enough
to avoid some numerical artefacts.  Pedagogical presentation are
given in \cite{ede00,deb08}.

\subsection{Formal specifications and proofs in computational geometry}

We work in the Calculus of
Inductive Constructions with Coq \cite{ber:cas,Coq}.
Related work on the description of geometric algorithms includes
\cite{pic:ber} also using Coq and \cite{mei:fle04} using Isabelle.  Concerning
graphs, \cite{bau:nip} gives a model of triangulations restricted to the
study of the five color theorem.  Hypermaps are also used intensively in
\cite{gon08} for the proof of the four-colour theorem.  A detailed comparison
is given in \cite{duf09}.  Hypermaps also play a significant role in the
formalization of packings by {\em tame graphs} in the proof of Kepler's conjecture \cite{obu:nip}.

Other work with close variants of the hypermaps used in this paper are
concerned with the formal study of geometric modelling \cite{pui:duf1},
surface classification \cite{deh:duf2}, image segmentation \cite{duf07}, and
a discrete form of the Jordan curve theorem~\cite{duf09}.

\section{Hypermaps}
\label{HY}

\subsection{Mathematical Aspects}
\label{MA}

\begin{figure}
\begin{center}
\includegraphics*[scale =.50]{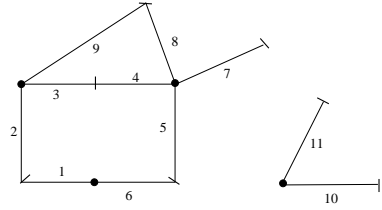}
\end{center}
\caption{An example of hypermap embedded on the plane.}
\label{fig:Dhmap1}
\end{figure}

\begin{definition}({\em Hypermap})
  \label{HD}\\
  (i) A {\em hypermap} is an algebraic structure $M = (D, \alpha_0,
  \alpha_1)$, where $D$ is a finite set, the elements of which are
  called {\em darts}, and $\alpha_0$, $\alpha_1$ are permutations on $D$.\\
\end{definition}
Intuitively, darts can be understood as half-edges, the permutation
\(\alpha_0\) usually connects the two darts of each edge, and
\(\alpha_1\) connects all the darts that meet on the same vertex of a
graph.  In general, the \(\alpha_0\) permutation could
link together an arbitrary number of elements, but in practice, it is
usually involutive.  Fig. \ref{fig:Dhmap1} gives an example of a
hypermap with
only three darts (darts 7, 10, and 11) that are not 0-linked to another one.
Such exotic darts may
always occur at intermediate stages during manipulation of maps.
For all other darts of Fig.~\ref{fig:Dhmap1},
the 0-successor of the 0-successor of each dart is the dart itself.

In Fig. \ref{fig:Dhmap1}, $\alpha_0$ and $\alpha_1$ are permutations
on $D = \{1,\ldots, 11\}$, then $M = (D, \alpha_0, \alpha_1)$ is a
hypermap. It is drawn on the plane by associating to each dart a
curved arc (here a simple line segment) oriented from a bullet to a
small stroke: $0$-linked (resp. $1$-linked) darts share the same small
stroke (resp. bullet). By convention, in the drawings of hypermaps on
surfaces, \(\alpha_k\) permutations turn {\em counterclockwise} around
strokes and bullets.

\subsection{Formal encoding}
\label{PS}
We use Coq's datatype declaration mechanism to define a two element
type {\tt dim} of dimensions and an infinite type {\tt dart} of darts, with a
special dart singled out for later purposes. This special dart is
called {\tt nil}. To describe embeddings we also add a type {\tt
  point} which is a pair of coordinates (real numbers).

Hypermaps are then described by collecting darts and links in a free map
linear data structure of type {\tt fmap}: 

\begin{verbatim}
Inductive fmap : Set :=
 V | I (m:fmap)(d:dart)(p:point) | L (m:fmap)(k:dim)(d1 d2:dart).
\end{verbatim}
This defines two operations, {\tt I} to add new dart {\tt d} in an
existing map {\tt m}, associating this dart with the location {\tt p},
and {\tt L} to add a link from dart {\tt d1} to dart {\tt d2} in
the map {\tt m}, at dimension {\tt k}.

This free data structure is too permissive: we
may add the same dart several times, we may link a dart that is not in
the map, etc.  We will see
later that hypermaps are free maps where some preconditions have been
verified before adding each dart and link, based on some helper functions.

A first function called {\tt exd} computes whether a given dart is
present in a map. Another pair of functions, named {\tt succ} and {\tt
  pred}, compute whether there is a link at a given dimension
with a given dart as source or target.  For each dimension, the
convention is to include in the free map only links that make up an
open path.  Thus, to represent a map where 
\(\alpha_k(d_1)=d_2\), \(\alpha_k(d_2)=d_3\) and \(\alpha_k(d_3)=d_1\), the
free map will only
contain a link from \(d_1\) to \(d_2\) and a link from \(d_2\) to
\(d_3\), or a link from \(d_2\) to \(d_3\) and a link from \(d_3\) to
\(d_1\), or a link from \(d_1\) to \(d_2\) and a link from \(d_3\) to
\(d_1\).  The \(\alpha_k\) functions are then
computed from the incomplete paths using a recursive algorithm that
simply traverses the free map structure.
  The formal notation in Coq syntax for the \(\alpha_k\)
functions of a given map \(m\) will be {\tt cA m k}. 

Hypermaps are then defined as free maps such that some preconditions
were verified before applying any of the {\tt I} or {\tt L}
operations.  The precondition {\tt prec\_I} for adding a dart in a
hypermap is that the dart should not already be present and should be
different from the special dart {\tt nil}. The precondition {\tt
  prec\_L} for adding a link is that the source and the target should
be darts in the map, the source should not already have a successor,
the target should not already have a predecessor, and the new link
should not be closing an open path.  As an example of our notations,
here is how our {\tt prec\_L} function is defined:
\begin{verbatim}
Definition prec_L (m:fmap)(k:dim)(x y:dart) :=
  exd m x /\ exd m y /\ ~succ m k x /\ ~pred m k y
    /\ cA m k x <> y.
\end{verbatim}
Verifying that a free map is indeed a hypermap can be described using
a simple recursive function {\tt inv\_hmap} that traverses the map and
verifies the preconditions at each step:

\begin{verbatim}
Fixpoint inv_hmap(m:fmap):Prop:=
  match m with
    V => True
    | I m0 x _ _ => inv_hmap m0 /\ prec_I m0 x
    | L m0 k0 x y => inv_hmap m0 /\ prec_L m0 k0 x y
  end.
\end{verbatim}

When {\tt m} is a hypermap, we prove that the \(\alpha_k\), or {\tt cA m k}, are permutations of the darts. Then, by construction, for every dart \(d\) the set \(\{d' | d' = \alpha^n_k(d)\}\) is finite and is called the {\em orbit} of \(d\)
at dimension \(k\). From the most abstract point of view, there is no difference between links at dimension 0 and links at dimension 1. However, to describe the subdivisions we are accustomed to manipulate, it will be better to ensure that orbits at dimension zero are edges, and thus contain only
two darts, while orbits at dimension one are vertices, and thus contain
only darts that are associated to the same geometrical point (see section \ref{Flipedge}).  We also
say that two darts \(x\) and \(y\) are in the same component if there
exists a path from \(x\) to \(y\) using the \(\alpha_k\) permutations at
each step.

When \(\alpha_0\) and \(\alpha_1\) are permutations, the composition of
their inverses \(\phi = \alpha_1^{-1} \circ \alpha_0^{-1}\) the orbits of
which are the {\em faces}.

Notions of components, paths, and orbits are independent from the
permutation being observed.  To handle all these in a regular fashion,
we developed a generic module.

Planar hypermaps can be characterised by counting their edges,
vertices, faces, and components \cite{duf08a}.  These remain topological
properties, independent from actual positions.

\begin{definition}({\em Euler characteristic, genus, planarity, Euler formula})\\
Let \(d\), \(e\), \(v\), \(f\), \(c\), be the numbers of darts, edges, vertices, faces, and components of a hypermap.\\
(i) The {\em Euler characteristic} of $M$ is $\chi= v + e + f - d$.\\
(ii) The {\em genus} of $M$ is $g = c -\chi / 2$.\\
(iii) When $g=0$, $M$ is said to be {\em planar}. 
It satisfies the {\em Euler formula}: $\chi = 2 * c$.
\end{definition}

Truly geometric aspects are described by observing the plane coordinates
associated to each dart in the {\tt I} operation.  Of course, embeddings are
consistent with the geometric intuition only if all darts in a vertex share
the same coordinates and the two darts that constitute an edge never have the
same coordinates.  An extra condition is that faces should not be too
twisted: we express this condition only for triangles, by stating that they
have to satisfy the {\em counter-clockwise} orientation predicate as already
used by Knuth in \cite{knu}.  

\begin{figure}
\begin{center}
\includegraphics*[scale =.35]{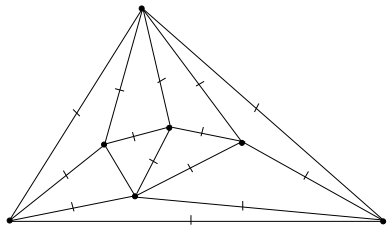}
\end{center}
\caption{A triangulation with triangular external face.}
\label{fig:triangul}
\end{figure}

In a nutshell, Knuth's orientation predicate relies on the existence 
of a 3-argument predicate on points (named {\tt ccw} in our development, Fig. \ref{fig:ccw}(a)) that satisfies
five axioms.  The first one expresses that if \(p,q,r\) satisfy {\tt ccw},
then so do \(q,r,p\) (in that order).  We shall also use a more complex
axiom, which we shall name Knuth's fifth axiom, with the following statement (Fig. \ref{fig:ccw}(b)):
\begin{verbatim}
Lemma axiom5 :
 forall p q r s t : point,
  ccw q p r -> ccw q p s -> ccw q p t -> 
    ccw q r s -> ccw q s t -> ccw q r t.
\end{verbatim}
\begin{figure}
\begin{center}
\includegraphics*[scale =.50]{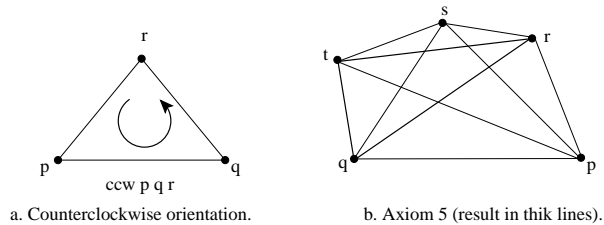}
\end{center}
\caption{Orientation of a triple of points (p, q, r) in the plane
 and the fifth axiom.}
\label{fig:ccw}
\end{figure}
Using all these concepts, we can state precisely what we mean by a
triangulation: a planar hypermap, where all edges have two darts,
and all faces have three vertices.
From the geometric point of view, this hypermap should also be
well-embedded: all edges contain darts with different geometric
locations, all triangles but one are oriented counter-clockwise.  The
one face that is not a counter-clockwise triangle correspond to the
external boundary.  In this first experiment, we have assumed this
external boundary to also be a triangle, but one that is oriented
clockwise (Fig. \ref{fig:triangul}).  
This simplification can also be found in well-known studies of the
Delaunay problem \cite{gs85}.  A hypermap that satisfies all these conditions
is said to be a {\em well-embedded triangulation}.  

\section{Split, Merge and Flip}
\label{FL}
In the previous sections, we have described the basic constructors of
hypermaps {\tt I} and {\tt L} and the many ways in
which we can observe maps and local parts of these maps.  Now, we will
study ways to transform maps.

\subsection{Splitting a k-orbit, merging two k-orbits}
\begin{figure}
\begin{center}
\includegraphics*[scale =.50]{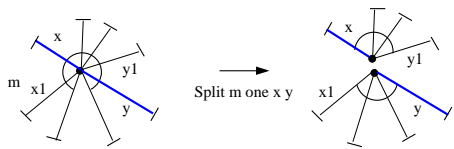}
\end{center}
\caption{Splitting a vertex.}
\label{fig:split2}
\end{figure}

When flipping edges, we need to detach darts from vertices (1-orbits).
A more general point of view is to consider that a vertex is actually
split into two parts while respecting the connection order.  To
understand the required transformations, we need to remember that links
are left open in the map structure. Before the split, one dart
has no 1-successor, after the split two of the darts taken from the split
vertex have no 1-successor.  The split operation is specified by stating the
two darts that have this property, let's assume these two darts are called
\(x\) and \(y\) (Fig. \ref{fig:split2}).

The split operation can be described for any dimension \(k\) and is
decomposed in two steps.  In the first step, one checks whether \(x\)
has a \(k\)-successor.  If it has one, then the darts \(z\) and \(t\)
in the \(k\)-orbit such that \(z\) has no \(k\)-successor and \(t\)
has no \(k\)-predecessor are computed, the \(k\)-link starting in
\(x\) is removed, and a link from \(z\) to \(t\) is added.  In this
step, the orbit is actually not changed, and we can call this operation
{\em shifting}.  In the second step, the one
link starting in \(y\) is removed.  The precondition for this
operation is that \(x\) and \(y\) should be different and in the same
\(k\)-orbit.  In our formal development this is described by a
function named {\tt Split} and we proved a few important properties of this operation, for instance that it preserves planarity and
commutativity with respect to \(x\) and \(y\).

\begin{figure}
\begin{center}
\includegraphics*[scale =.50]{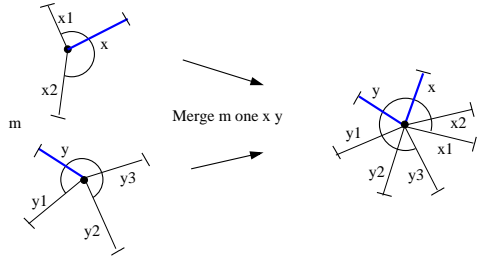}
\end{center}
\caption{Merging two vertices.}
\label{fig:mergedef2}
\end{figure}

To merge two orbits, we need to choose a dart \(x\) in one of the
orbit and a dart \(y\) in the other, with the intention that the
\(k\)-successor of \(x\) will be \(y\) in the new map (Fig. \ref{fig:mergedef2}).  Of course, a
first step is to make sure that the two orbits are shifted in such a way
that \(x\) has no successor and \(y\) has no predecessor before adding
a link from \(x\) to \(y\). This operation has a pre-condition imposing that \(x\) and \(y\) are not in the same orbit.  When considering merging at dimension 1 (merging vertices), the effect on edges and vertices is quite obvious, but less clear for faces \cite{duf08a,duf09b}.

\subsection{Flipping an edge}
\label{Flipedge}

Flipping an edge actually consists in first removing an edge thus
``merging'' two adjacent triangles, and then adding back a new edge
between two different vertices from the merged face.  Actually, the
two vertices between which a new edge is added are neighbors to the
two vertices from which the first edge was removed.  The number
of darts in the map is preserved, so that the edge that is removed in
the first step can be viewed as moved from a pair of vertices to
another one.  The first step of removing an edge is described using two
split operations, while the second step of adding back a new edge is
described using two merge operations.  Embeddings must then be updated
to respect the requirement that all darts in a vertex share the same
location.

\begin{figure}
\begin{center}
\includegraphics*[scale =.50]{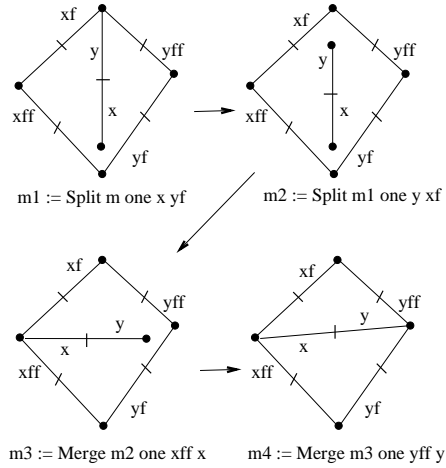}
\end{center}
\caption{Four topological steps of Flip.}
\label{fig:flip}
\end{figure}

The topological steps are illustrated in Fig. \ref{fig:flip}.  
The precondition for this operation is that the two darts in the edge should
be in different faces and connected to vertices of 3 darts or more.

In intermediate steps, the subdivision is no longer a
triangulation: the merged face has a different number of vertices, the
detached edge is a component of its own, etc.  However, we describe a pair
of preconditions, named {\tt prec\_Flip} and {\tt prec\_Flip\_emb}
that ensure that the flipping operation as a whole preserves the important
topological properties, for instance planarity, having only two-dart edges and
three-vertex faces and the embedding properties, for instance that all
darts in a 1-orbit (a vertex) share the same coordinates and that
all triangles but for the external face are oriented counter-clockwise.
The precondition for topological properties ({\tt prec\_Flip})
is that the flipped edge consists
of darts belonging to distinct faces and to vertices with at least three 
darts. The pre-condition for embedding properties ({\tt prec\_Flip\_emb})
is that the four
points in the intermediate merged face should constitute a convex quadrangle.
In our formal development, we actually prove that {\tt prec\_Flip} is
sufficient to preserve the important topological properties, that the
{\tt prec\_Flip\_emb} is sufficient to preserve the well-embedding properties,
and that {\tt prec\_Flip\_emp} implies {\tt prec\_Flip} \cite{duf09b}.
We shall see that our algorithm for Delaunay triangulation only requires
flipping edges that satisfy these predicates.

\section{The Delaunay Criterion}
\label{DP}
A triangulation satisfies the Delaunay criterion when none of the
vertices occurs inside the circumcircle of a face.  In other words, there
are no illegal edges.  In our development we defined a four-argument predicate
 {\tt in\_circle} to express that a point is inside the circumcircle of
three other points.

\begin{definition}({\em Illegal edge}) 
\label{Il}\\
An edge is {\em illegal} in a well-embedded plane triangulation when:\\
(i) its two adjacent triangles are counterclockwise oriented 
(which excludes the external face); \\
(ii) the vertex of one of the two triangles which is not an extremity of the
edge is inside the circumcircle of the other triangle.
\end{definition}

\begin{figure}
\begin{center}
\includegraphics*[scale =.45]{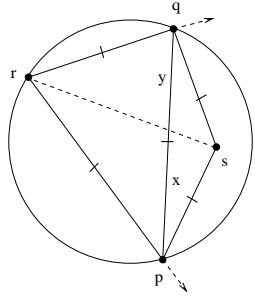}
\end{center}
\caption{Point $s$ is in the circumcircle of $(p,q,r)$}
\label{fig:legal}
\end{figure}

This notion is illustrated in Fig. \ref{fig:legal}, where \(s\)
is inside the circumcircle of triangle \((p, q, r)\), at the right
of \(pq\).  Note that this property is symmetrical with respect to
the two triangles.  When an illegal edge is detected, we know that the
preconditions for the flip operation are satisfied.  When the operation is performed, the new edge produced by this flip operation is
legal.  This contains two parts: the two new triangles are oriented, and
the circumcircles of each new triangle does not contain the fourth point.

More precisely, the important property, called {\tt exchange} in our
formal development, asserts that when two adjacent triangles $(p,q,r)$
and $(q,p,s)$ are oriented counterclockwise and $s$ is in the
circumcircle of $(p,q,r)$, then the triangles $(r,s,q)$ and $(s,r,p)$
are also oriented counterclockwise (Fig.\ref{fig:legal}).

Proving this part required some effort.  We actually showed that, when
\(p\), \(q\), \(r\), and \(s\) are in the conditions of the lemma,
then there exists a fifth point \(t\) so that \(p\), \(t\) \(q\), \(r\),
and \(s\) are in the conditions of Knuth's fifth axiom for the 
orientation predicate.  This point is simply the one obtained by rotating
the center of the circumcircle by a quarter-turn around \(p\).  We can then
use Knuth's fifth axiom to conclude that \(p,s,r\) is oriented counterclockwise.
A symmetric proof (with a rotation around \(q\)) yields that \(q,r,s\) is
oriented counterclockwise.  This symmetric proof is implemented by
copying and pasting the formal development, {\sl mutatis mutandi}.  Uses
of Knuth's first axiom then yield the result.

The 3-argument predicate {\tt ccw} is computed from
point coordinates through a simple determinant:
\[\left|\begin{array}{ccc}~x_p~&~y_p~&~1~\\x_q&y_q&1\\x_r&y_r&1\end{array}\right|\]
The boolean condition is represented by the sign of the determinant and the
condition of degeneracy, that three points are never aligned, ensures that
this determinant is non-zero.  The 4-argument predicate, {\tt in\_circle}
is also computed through the sign of a simple determinant:
\[\left|\begin{array}{cccc}~x_p~&~y_p~&~{x_p}^2 + {y_p}^2~& ~1~\\
  x_q&y_q&{x_q}^2 + {y_q}^2& 1\\
  x_r&y_r&{x_r}^2 + {y_r}^2& 1\\
  x_s&y_s&{x_s}^2 + {y_s}^2& 1\end{array}\right|\] 
Knuth's five axioms are easily proved using algebraic tools (in Coq,
mostly the {\tt ring} tactic) from these analytic definitions
\cite{knu,pic:ber}.  Proving the existence of the point \(t\) a few
paragraphs above actually relies on a stronger tool, a tactic called
{\tt psatz} (the name comes from {\em positivstellensatz})
and able to handle simple cases of non-linear formulas,
available only in recent versions of Coq \cite{bes06}.

\section{Termination based on finiteness}
\label{TF}
Traditional approaches to ensure the termination of algorithms rely on
structural recursion for the simplest algorithms and well-founded orders for
the others.  In this work, we took the novel approach of relying on three
features:
\begin{itemize}
\item We rely on the fact that the number of triangulations embedded on a
  given finite set of points and using a finite set of darts is finite,
\item We exhibit an order on triangulations that is not well-founded,
  but we show that flipping an illegal edge implies a strict decrease
  in that order,
\item We then rely on the fact that any transitive, irreflexive, and
antisymmetric relation \(R\) is well-founded when restricted on a finite set.
\end{itemize}

\subsection{A generic library for finiteness}
For the formal development, we describe a minimal description of finiteness for subsets of a type.  First, we represent each subset of a type {\tt T} by a predicate on {\tt T}, i.e., a function of type {\tt T -> Prop}.  Then we express finiteness by stating that all elements satisfying the predicate are found in a list. This is specified by the following datatype declaration:

\begin{verbatim}
  Record fset (T:Type) := mkfs {
    prd :> T -> Prop;
    fs_enum : list T;
    _ : forall x, prd x -> In x fs_enum
  }.
\end{verbatim}
This declaration states that a finite set on type {\tt T} is described
by the characteristic predicate {\tt prd} of type {\tt T -> Prop} and
a list {\tt fs\_enum} which enumerates the elements that satisfy {\tt
  prd}.  Actually our definition is quite lenient, because it makes it
possible to have in the list more elements than those satisfying the
predicate.  The list is very useful because it gives a simple way to
iterate over all the elements in the finite set (and with
our lenient definition, risking to see several times the
same elements and elements outside the set).  This
method, of associating two points of view (predicate or covering list)
over a simple notion (finite set) is directly inspired from the approach
to describe finite sets in the {\tt ssreflect} package \cite{GMRTT:GROUPS07}.

We then show that finiteness is preserved by cartesian product, 
disjoint sum, inclusion, inverse image through an injection, construction
of lists of fixed length, construction of lists of bounded length,
and construction of lists without duplication.

To show that the triangulations we consider are in a finite set, we
start by computing from any map the list of darts and the set of points
that appear in this map.  We show that this list of darts and this set of
points is preserved during flips.  It is easy for the list of darts because
the order of the {\tt I} constructors in the {\tt fmap} structure is not
modified by the basic shift, split, or merge operations.  For points, it is
harder, because a flip operation changes the number of darts that use
a given coordinate and we need to show that the set is preserved modulo
a possible change in the order and number of times each point is inserted.
We do this by defining a sorting function with removal of duplicates (an
insertion sort algorithm with an extra test to detect duplications) and applying
this sorting function on the list of points used in the triangulation.  We then
show that the list of points obtained after a flip operation, once sorted and
cleaned from duplicates, is preserved through flipping.

We then show that all maps built on the same list of darts and the same
set of points are in a finite set, obtained using cartesian products, sums,
etc.
\subsection{A strict order on triangulations}

As a complement to the finiteness property, we must exhibit a strict order
that decreases every time an illegal edge is flipped.

It is well known that Delaunay triangulation is closely related to
computing the three-dimensional convex hull of points projected from
the horizontal plane to the revolution paraboloid with equation \(z =
x^2 + y^2\).

Given four points \(p\), \(q\), \(r\), and \(s\) in a
three-dimensional space, the determinant obtained from their
coordinates by adding a column of ones actually computes a value which
is proportional to the volume of the tetrahedron defined by these four
points.
\[\left|\begin{array}{cccc}~x_p~&~y_p~&~z_p~& ~1~\\
  x_q&y_q&z_q& 1\\
  x_r&y_r&z_r& 1\\
  x_s&y_s&z_s& 1\end{array}\right|\] 

Thus, the determinant computed in Section~\ref{DP} to decide whether a
point occurs inside the circumcircle of a triangle actually computes
the volume of the tetrahedron defined by the four projections of the
points from the plane to the paraboloid. When considering two adjacent
triangles and the triangles obtained after flipping the common edge,
we can compute the volume between these two triangles
and the corresponding triangles using the projected points in the
paraboloid. The two configurations yield two different volumes.
The difference of volume is exactly the volume of the
tetrahedron based on the points in the paraboloid, and it is positive
when the projected triangles switch from a concave position to a
convex one.

To compute each individual volume, we decompose the prism-like shape
into three tetrahedra, each being computed using a
determinant. Showing the relation between the volumes of the two
prism-like shapes before and after the flip operation and the
determinant used for the {\tt in\_circle} predicate is an easy task
using Coq's ring tool.

To compute the accumulated volume, we simply enumerate the edges of the
map and add the triangle obtained as the \(\phi\)-orbit for each edge.
Of course, each triangle is thus represented three times, but this
does not matter for our decreasing argument.  We simply need to show that the
volume computed only changes for the six darts whose \(\phi\)-orbit
 changes during the flip operation.

\subsection{Describing a terminating function}
To describe a terminating function, we rely on a type {\tt tri\_map}, which
combines a free map and the proof that it is a well-embedded triangulation.
This type is defined as a conventional Coq sigma-type:
\begin{verbatim}
Definition tri_map := {m | inv_Triangulation m /\ isWellembed m}.
\end{verbatim}
The natural projection returning the free map is written
{\tt p\_tri}.

We then define a function {\tt step\_tri}, from type {\tt tri\_map} to itself,
which performs a flip when the map contains an illegal edge.  This
function relies on the proofs that 
flip preserves the property of being a well-embedded triangulation.
We also define a {\tt final\_dec} function that detects when
 there are no illegal edges.

Last we define a function {\tt nat\_measure} which first constructs 
the final set of all triangulations using
the same darts and points, with its enumerating list and then counts the
triangulations in this list whose volume is smaller than the current one.
This natural number decreases at every flip on a triangulation
that contains an illegal edge, i.e., every derivation that does not 
satisfy the {\tt final} predicate.

The recursive algorithm is not structural recursive, so we need to use
one of the tools provided in the Coq system to support general forms
of recursion.  Here, we use the {\tt Function} command, which accepts a
definition as long as one can prove that some measure (a natural
number) decreases at each recursive call.  We first prove the lemma
{\tt non\_final\_step\_decrease} and then provide it to the {\tt
  Function} command.
\begin{verbatim}
Lemma non_final_step_decrease :
  forall m, ~final (p_tri m) ->
   (nat_measure (step_tri m) < nat_measure m)%nat.
...

Function delaunay' (t : tri_map) {measure nat_measure} :=
 if final_dec (p_tri t) then
   (p_tri t)
 else
   delaunay' (step_tri t).
\end{verbatim}
Computing the finite set of all triangulations is expensive (an
exponential cost in the number of darts and points), but this
computation is not actually done in the algorithm, it is used as a
logical argument for termination.  This computation is actually
removed from the derived code produced by Coq's extraction facility.
\section{Solving the Delaunay problem} 
\label{DS}
It only makes sense to run the algorithm on well-embedded triangulations.
Thus, our {\tt Delaunay} function takes as argument a map and the proofs
that this map is a triangulation and that it is well-embedded.  It then calls
the {\tt delaunay'} function with the adequate element of type {\tt tri\_map}.
\begin{verbatim}
Definition Delaunay (m : fmap)(IT inv_Triangulation m)
    (WE:isWellembed m) : fmap :=
  delaunay' (exist _ m (conj IT WE)).
\end{verbatim}

In our formal proof, we show that the end result of the {\tt Delaunay}
function returns a well-embedded triangulation that contains no illegal
edges.  For instance, we have the following statement:

\begin{verbatim}
Theorem no_dart_illegal_Delaunay : 
  forall (m : fmap)(IT: inv_Triangulation m)(WE: isWellembed m),  
      no_dart_illegal (Delaunay m IT WE). 
\end{verbatim}
In English words, we quantify over all free maps that satisfy two
predicates.  The first predicate {\tt inv\_Triangulation} captures all
the conditions for the map to be a correct triangulation in the
topological sense: it is a correct hypermap, 0-orbits have two
elements only, faces have three elements.  The second predicate
{\tt isWellembedded}
expresses that the coordinates are consistent: all darts in the same point
share the same coordinates, all triangles are oriented.  The hypotheses
that the map satisfies these predicates are given names {\tt IT} and {\tt WE}
respectively.  The function {\tt Delaunay}
that computes the Delaunay triangulation takes
these hypotheses as arguments.  We then use a predicate {\tt no\_dart\_illegal}
to express that the Delaunay condition is always satisfied: it is never the
case that the extra vertex of an adjacent triangle is inside the circumcircle
of a given triangle.

\section{Conclusion}
\label{CL}
The one missing element of this algorithm is a starter: given an arbitrary set
of points inside a triangle, we need to produce the initial triangulation.
Developing a naive algorithm, with only the requirement that the triangulation
should be well-formed, should be an easy task.  Actually, if the three points
describing the external face are given first, an possible algorithm is
a simple structural recursive function on the list of points.

All numeric computations are described using ``abstract perfect'' real
numbers.  In practice, specialists in algorithmic geometry know that numeric
computation with floating point numbers can incur failures of the algorithm
by failing to detect illegal edges, or by giving inconsistent results for
several related computations \cite{yp06,ket08}.  For instance, rounding errors
could make that
both an edge and its flipped counterpart could appear to be illegal, thus
leading to looping computation that is not predicted by our ideal formal model.
However, we know that all predicates are based on determinant computations,
hence polynomial computation, and it is thus sufficient to ensure that
intermediate computations are done with a precision sufficiently higher than
the precision of the initial data to guarantee the absence of errors
introduced by rounding.  Thus, the ``theoretical'' correctness of the algorithm
can be preserved in a ``practical'' sense if one relies on a suitable
approach to increase the precision of numeric computations, as in \cite{mel:pio,pri91,Fou07}.

Our whole development from the hypermap specifications and proofs
up to the Delaunay properties reaches about $70,000$ Coq lines, with
more than $300$ definitions and $700$ lemmas and theorems.  Thanks to
the {\sl extraction} facility provided in the Coq sytem, an Ocaml version
of the algorithm can be obtained (where every computation on real numbers
is replaced by computation on unbound integers for instance, since
division is never used in the algorithm) \cite{duf:ber10}.

We described the most naive algorithm for the Delaunay problem.  We
believe that most of the framework concerning the topology will be
re-usable when studying other algorithms for this problem
\cite{gs85,ede00,deb08}.  Also, our proof reason on abstract models
given as Coq programs, not actual programs
designed for efficiency.  Previous experiments in the formalization of
efficient algorithms \cite{BMZ02} show that the proofs at an abstract level are
a useful first step for the study of efficient programs given in
an imperative language.

Our framework is a sound basis for subsequent software developments
with triangulations and Flip in computational geometry and geometric
modeling, for instance in the way of
\cite{ber:duf,duf:pui,duf07,bru:duf:mag09} where hypermaps are
represented by linked lists.  The functional, side-effect-free
approach in this formal description has been very useful for the proofs.
However, for efficiency purpose it is crucial to relate this functional description with imperative implementations.
\\ \\
\noindent{\bf Acknowledgments.}  We wish to thank L. Pottier,
T.-M. Pham, and S. Pion for their suggestions in establishing some of
the geometric proofs.

\end{document}